\documentclass[nolinenumbers,twocolumn]{aastex631}
\usepackage{diagbox} % 引入斜线宏包
\usepackage{diagbox}
\usepackage{multirow}
\usepackage{array}
\usepackage{comment}
\usepackage{amsmath}

\DeclareUnicodeCharacter{02BC}{<definition>}
\DeclareUnicodeCharacter{FF1A}{<definition>}
\begin{document}

\title{X-ray Binaries: a potential dominant contributor to the cosmic ray spectral knee structure}

\correspondingauthor{Jianli Zhang}
\email{jlzhang@bao.ac.cn}
\correspondingauthor{Yuhai Ge}
\email{geyh@ihep.ac.cn}
\correspondingauthor{Lin Nie}
\email{nielin@ihep.ac.cn}

\author{Hua Yue}
\affiliation{State Key Laboratory of Particle Astrophysics, Institute of High Energy Physics, Chinese Academy of Sciences, 100049 Beijing, China}
\affiliation{School of Physical Sciences, University of Chinese Academy of Sciences, No.19(A) Yuquan Road, Beijing, China}

\author{Jianli Zhang$^{\dag}$}
\affiliation{National Astronomical Observatories, Chinese Academy of Sciences, 100101 Beijing, China}
%jlzhang@bao.ac.cn}
\author{Yuhai Ge$^{\dag}$}
\affiliation{State Key Laboratory of Particle Astrophysics, Institute of High Energy Physics, Chinese Academy of Sciences, 100049 Beijing, China}
\author{Lin Nie$^{\dag}$}
\affiliation{School of Physical Science and Technology, Southwest Jiaotong University, Chengdu, 610031, China}
\affiliation{State Key Laboratory of Particle Astrophysics, Institute of High Energy Physics, Chinese Academy of Sciences, 100049 Beijing, China}

\author{Peipei Zhang}
\affiliation{Intelligence and Information Engineering college, Tangshan University, Hebei, 063000, China}
\author{Wei Liu}
\affiliation{State Key Laboratory of Particle Astrophysics, Institute of High Energy Physics, Chinese Academy of Sciences, 100049 Beijing, China}
\affiliation{TIANFU Cosmic Ray Research Center, Chengdu, Sichuan, China}
\author{Yiqing Guo}
\affiliation{State Key Laboratory of Particle Astrophysics, Institute of High Energy Physics, Chinese Academy of Sciences, 100049 Beijing, China}
\affiliation{School of Physical Sciences, University of Chinese Academy of Sciences, No.19(A) Yuquan Road, Beijing, China}
\affiliation{TIANFU Cosmic Ray Research Center, Chengdu, Sichuan, China}
\author{Hongbo Hu}
\affiliation{State Key Laboratory of Particle Astrophysics, Institute of High Energy Physics, Chinese Academy of Sciences, 100049 Beijing, China}
\affiliation{School of Physical Sciences, University of Chinese Academy of Sciences, No.19(A) Yuquan Road, Beijing, China}
\affiliation{TIANFU Cosmic Ray Research Center, Chengdu, Sichuan, China}
\begin{abstract}
``PeVatrons" refer to astrophysical sources capable of accelerating particles to energies $\sim$PeV and higher, potentially contributing to the cosmic ray spectrum in the knee region. Recently, HAWC and LHAASO have discovered a new type PeVatrons -- X-ray binaries, allowing us to investigate in greater depth of the contributions of these sources to cosmic rays around the knee region. There are hundreds of X-ray binaries in our galaxy observed, which are potential PeVatrons.
In this work, we derive the radial distribution of X-ray binaries in the Galaxy. Then we use the DRAGON package to calculate energy spectrum, anisotropy of cosmic rays as well as the resulting diffuse gamma ray emissions, after considering them as factories of cosmic rays in the knee energy bands. Our findings show that the contributions from X-ray binary PeVatrons may be dominant. More X-ray binary PeVatrons can be observed by LHAASO and HAWC in the future, and will confirm the contribution of X-ray binaries to high energy cosmic rays.

\end{abstract}


\section{Introduction} \label{sec:intro}
The origin of high energy cosmic rays (CRs) is a long standing unresolved issue in particle astrophysics. The energy spectrum of CRs can be roughly described by a power-law spectrum of $\rm E^{-2.7}$, extending up to the knee region at around 3 PeV~\citep{2024PhRvL.132m1002C}, beyond which the spectrum softens~\citep{1959bSov.Phys.JETP...35..441}. CRs below the knee region are believed to be produced and accelerated within the Milky Way~\citep{1961PThPS..20....1G}, indicating the presence of PeV acceleration sources within our galaxy, known as PeVatrons~\citep{2021Natur.594...33C,2024ApJS..271...25C}. So uncovering the galactic acceleration limit and identifying CR sources is the key to investigate of the knee region in CR spectra.

Currently, it is widely believed that supernova remnants (SNRs) are the primary acceleration sources of Galactic CRs. This is attributed to the nonlinear diffusive shock acceleration mechanisms that may occur in SNRs, which can accelerate CR particles to the PeV energy range ~\citep{2010ApJ...718...31P,2010ApJ...708..965Z,2023ApJ...952..100N}. Additionally, the radiation they produce successfully explains the multi-wavelength non-thermal emission observed from SNRs ~\citep{2012APh....39...12Z}. Furthermore, the so-called ``standard propagation model" of CRs considers SNRs as the primary sources of Galactic CRs, which partially explains the observed CR and diffuse gamma-ray emissions from the Milky Way. However, other celestial objects, such as pulsars, pulsar wind nebulae (PWNe), ``super-bubbles", the Galactic center, and X-ray binaries (XRBs), are also commonly considered as candidate sources of Galactic CRs. 
In recent years, the research in the origin of very high energy CRs has changed significantly due to the large detection area and long duty cycle of the LHAASO (Large High Altitude Air Shower Observatory)~\citep{2010ChPhC..34..249C}, which has identified 12 ultra high energy (UHE) gamma-ray sources, with the highest energy reaching 1.4 PeV~\citep{2021Natur.594...33C}. Recently, LHAASO released its first catalog, discovering dozens of PeVatrons throughout the Milky Way, including but not limited to PWNe, Binaries, and Star forming regions~\citep{2024ApJS..271...25C}. Especially, HAWC discovers the Ultra-high-energy gamma-ray bubble around microquasar V4641 Sgr~\citep{2024Natur.634..557A_HAWC_V4641} (A microquasar is an XRB system which launches and collimates relativistic jets), and LHAASO discovers Ultra-high-Energy Gamma-ray Emission from several X-ray binaries with Black Hole Jet system~\citep{2024arXiv241008988L_BLJet}. These findings from LHAASO provide a rich and crucial set of candidate samples for exploring the origins of PeV CRs within our galaxy. It seems to be that the CRs around the knee may be contributed by the XRBs.

It is worth noting that, on the other hand, only a few SNRs have been observed to provide evidence for CR particle acceleration to sub-PeV energies. For instance, the energy spectra of the SNRs IC443 and W44, observed by the Fermi Gamma-ray Space Telescope, exhibit the characteristic "bump" in the 70 MeV energy range, indicative of $\rm \pi_0$ meson decay ~\citep{2011ApJ...742L..30G,2013Sci...339..807A}. Meanwhile, with the ongoing accumulation of observational data from high-energy instruments like LHAASO, more and more binary systems have been detected emitting electromagnetic radiation beyond the PeV energy range. For example, among the 12 microquasars recently reported by LHAASO, 7 have been observed to emit radiation above 100 TeV~\citep{2018Natur.562...82A,2024arXiv241008988L}. In the case of SS433, extended radiation observed in the central region is spatially correlated with the surrounding gas cloud, suggesting hadronic origins for the radiation.

Binary systems have long been believed to be accelerators of Galactic CRs, converting gravitational potential energy from accreted material or the rotational energy of compact objects into the energy of CR particles, thereby accelerating them ~\citep{2024JHEAp..43...93K,2020ApJ...889..146S,1998NewAR..42..579A}. This suggests that binary systems may be one of the major contributors to CRs near the PeV energy range.

With the deployment of new-generation ~\citep{2024ChPhC..48f5001C}, higher-sensitivity observatories capable of probing higher energy ranges, the origins of the complex spectral structure in the CR PeV region are expected to be clarified.
The ``knee" structure of the CR spectrum, first observed 60 years ago in the PeV energy range, is characterized by a clear inflection point in the energy spectrum. The exact cause of the ``knee" structure remains uncertain. Various hypotheses, including those regarding CR acceleration~\citep{1999JETP...89..391B,2010ApJ...718...31P,2004APh....21..241H,2003APh....19..193H} and propagation origins~\citep{2001NuPhS..97..267L,2003ICRC....1..315O}, as well as nearby source origins~\citep{2021BLPI...48...31E}, have been proposed to explain this spectral feature, but no unified theoretical explanation has been established. To resolve the cause of the ``knee" structure, high-energy and highly accurate observational data in the PeV range is crucial. Extremely high-energy electromagnetic radiation, which is a secondary product of CR acceleration and propagation, directly provides evidence of the presence of PeV CRs in the local environment.

This paper focuses on the recent continuous observations of PeV radiation from microquasars by LHAASO, statistically analyzes the distribution of the observed XRBs within the Milky Way, and explores the role of XRBs as contributors to CRs by embedding them into the CR propagation model. The study aims to investigate the XRB origin of the CR ``knee" spectral structure.
The paper is organized as follows: in Section \ref{sec:X-Ray_binay_distribution}, we describe XRBs and Sample processing methods. In Sections \ref{sec:method} and \ref{sec:results}, we describe the model and the results. Finally, in Section \ref{sec:summary}, a concluding discussion and an outlook are given.

\section{The Distribution of X-Ray binaries} \label{sec:X-Ray_binay_distribution}

\subsection{The Observation of X-Ray binaries} \label{subsec:X-Ray_binaries}

XRBs emit bright light in the sky when observed in X-rays, making them one of the brightest objects in the electromagnetic spectrum.

XRBs, such as those studied by ~\citep{verbunt1993origin} and ~\citep{2017hsn..book.1499C}, represent a class of binary stellar systems where a compact object, potentially a stellar-mass black hole or neutron star known as the accretor, orbits a common center of gravity with a companion star providing material. These binaries can be categorized into two main groups based on the mass of the companion star.

High-mass X-ray binaries (HMXBs) studied by ~\citep{2012MNRAS.425..595R}, and ~\citep{2019NewAR..8601546K} feature massive companion stars which  shed material through stellar winds in an ongoing manner. This material is then captured and heated by the central compact object, leading to the emission of non-thermal X-rays.

Now, 5 HMXBs have been detected at high energies by Fermi-LAT, including LS 5039, Cygnus X-1, Cygnus X-3, SS 433 and LS I+61 303. Among them, LS 5039 and LS I+61 303 are classified as gamma-ray binaries due to their prominent high-energy gamma-ray emissions, while the others are standard HMXBs.
At least three are detected in Ultra-high-Energies by LHAASO, including SS 433, Cygnus X-3 and Cygnus X-1.

Relatively speaking, low-massive X-ray binary stars (LMXBs) have a small companion star and a black hole or neutron star undergoing accretion. These compact celestial bodies play a crucial role in the initiation of strong jets, which generate non-thermal radiation across various wavelengths ~\citep{2022abn..book.....C}.

To date, one LMXB (GRS 1915+105~\citep{2001A&A...373L..37G}) has been conclusively detected at high energies~\citep{2024arXiv241010396M}. But in Ultra-high-Energy Gamma-ray, three are confirmly detected by LHAASO and HAWC, V4641 Sgr, GRS 1915+105 and MAXI J1820+070~\citep{2020ApJ...893L..37T}, especially for V4641 Sgr, the energy extened to 1 PeV.

Many recent studies have suggested that stellar mass black holes (BHs; therefore also known as BHXBs) in X-ray binary stars (XRBs) are proposed as additional candidates for CR sources (~\citep{2005ChJAS...5..183T},~\citep{2020MNRAS.493.3212C},~\citep{2023MNRAS.524.1326K}). In particular, stellar mass black holes in XRBs exhibit relativistic jets, accreting material from their companion stars (~\citep{2004MNRAS.355.1105F}; ~\citep{2006ARA&A..44...49R}). They are efficient particle accelerators to produce non-thermal emission up to PeV~\citep{2022A&A...665A.145E}. Other types of XRBs with coronal emission can also act as CR sources~\citep{2024ApJ...975L..35F}.

The presence of X-ray binary systems (XRBs) within our galaxy holds significant importance. According to Cooper's analysis~\citep{2020MNRAS.493.3212C}, which is based on population synthesis findings from ~\citep{2020A&A...638A..94O}, it is estimated that there are approximately 5531 XRBs in the Milky Way. Recent X-ray studies conducted at the Galactic center indicate the potential existence of hundreds to thousands of black hole X-ray binaries (BHXBs) in the parsec-scale vicinity of the Galactic center (~\citep{2018Natur.556...70H}; ~\citep{2021ApJ...921..148M}). Referring to these observations, Cooper ~\citep{2020MNRAS.493.3212C} provided estimation suggesting a few thousand XRBs are likely present in the galaxy. These XRBs could play a significant role in shaping the CR spectrum and potentially have a dominant influence slightly above the ``knee" region before the transition to extragalactic sources.

 Recently, ~\citep{2023A&A...677A.134N} and ~\citep{2023A&A...675A.199A}  present a new catalogue of LMXBs and HMXBs in the Galaxy. There are 172 HMXBs and 360 LMXBs in the catalogue. This catalogue was selected for its homogeneity and X-ray detections from Swift/XRT, Chandra, and XMM-Newton. Based on this catalogue, we can produce the distribution of the LMXBs and HMXBs in the Galaxy. 

 Since most microquasars observed by LHAASO are {either} HMXBs {or LMXBs}, it is therefore reasonable to approximate their spatial distribution {by combining those of HMXB and LMXB populations}, especially considering the limited number of presently known microquasar sources.

\begin{figure*}[ht!]
\centering
\includegraphics[width=0.8\linewidth]{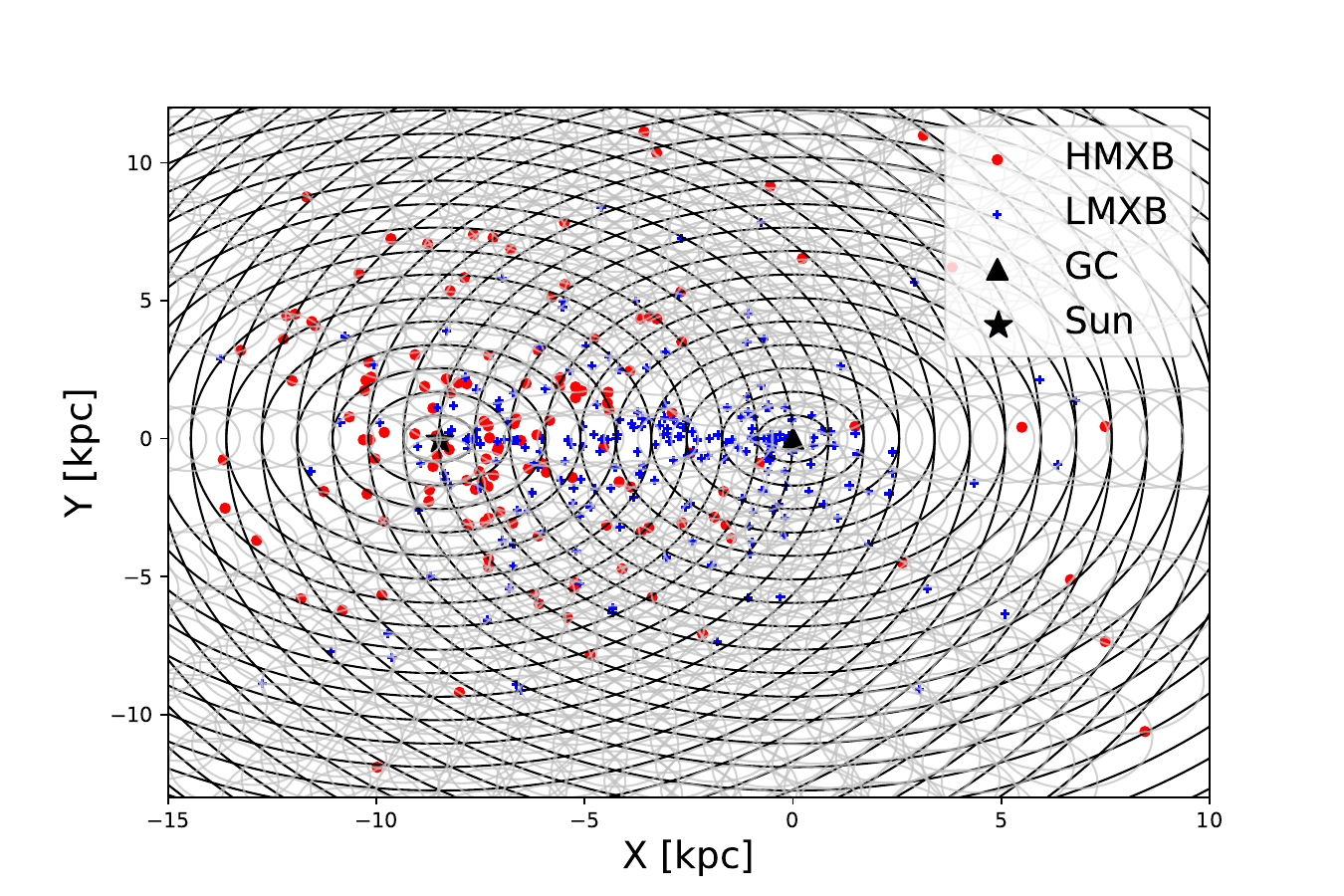}
\caption{Projection of the sample X-ray binaries (XRBs) on to the Galactic plane. High-mass X-ray binaries (HMXBs) and low-mass X-ray binaries (LMXBs) are marked as red dots and blue pluses, respectively. The GC is the origin of the coordinate system and the coordinate of the Sun is (-8.5, 0.0). Other lines and symbols are described in the context.}
\label{projection}
\end{figure*}

172 HMXBs and 360 LMXBs in the catalogue are projected onto the Galactic plane, as depicted in Fig~\ref{projection}. We adopt the traditional distance from the Sun to the Galactic Center as $\rm R_{\odot} = 8.5$ kpc, although more accurate observations ~\citep{brunthaler2011bar} suggest this distance is $\rm R_{\odot} = 8.3 \pm 0.23 $ kpc. Therefore, in our Cartesian coordinate system, the Galactic Center is located at the origin (0, 0) marked by a triangle in Fig~\ref{projection}, while the Sun is positioned at (-8.5, 0) marked by a pentagram.

\subsection{Selection and Processing Methods} \label{subsec:results_LHAASO}
HMXBs and LMXBs essentially trace different stellar populations (young vs. old). However, according to recent research~\citep{2021PASJ...73.1315I}, when parameters of the Milky Way are substituted into the constructed XRB population model, the ratio of HMXBs to LMXBs approaches 1:1. This initial study prioritizes evaluating XRBs' aggregate CR contribution over subtype differences, so we treat HMXBs and LMXBs as a single population.

We outline the selection criteria and data processing techniques applied to the analysis. The samples analyzed here are exclusively X-ray-selected XRBs, drawn from catalogs by ~\citep{2023A&A...677A.134N} and ~\citep{2023A&A...675A.199A}. These catalogs compile sources detected via X-ray surveys (Swift/XRT, Chandra, XMM-Newton) with uniform flux thresholds, explicitly excluding radio-selected samples. To correct for distance-dependent selection effects, we utilize a gridding method similar to those previously employed in the literature ~\citep{yusifov2004revisiting, xie2024modeling}, from which the pulsar distribution is derived.

In this study, we construct a quasi-regular grid on the Galactic plane by drawing concentric circles with equal radial spacing around both the Sun and the Galactic Center (GC). The radii of these circles, denoted as $\rm r_{i}$ for the Sun and $\rm R_{j}$ for the GC, are chosen such that the radial distance between adjacent circles is $\rm \Delta R=R/10=0.85\rm~kpc$, ensuring a sufficient number of intervals along the Sun-GC axis. The intersections of these circles define a grid, with each grid cell corresponding to a circular region centered at the intersection points.

\begin{figure}[ht!]
\centering
\includegraphics[width=\columnwidth]{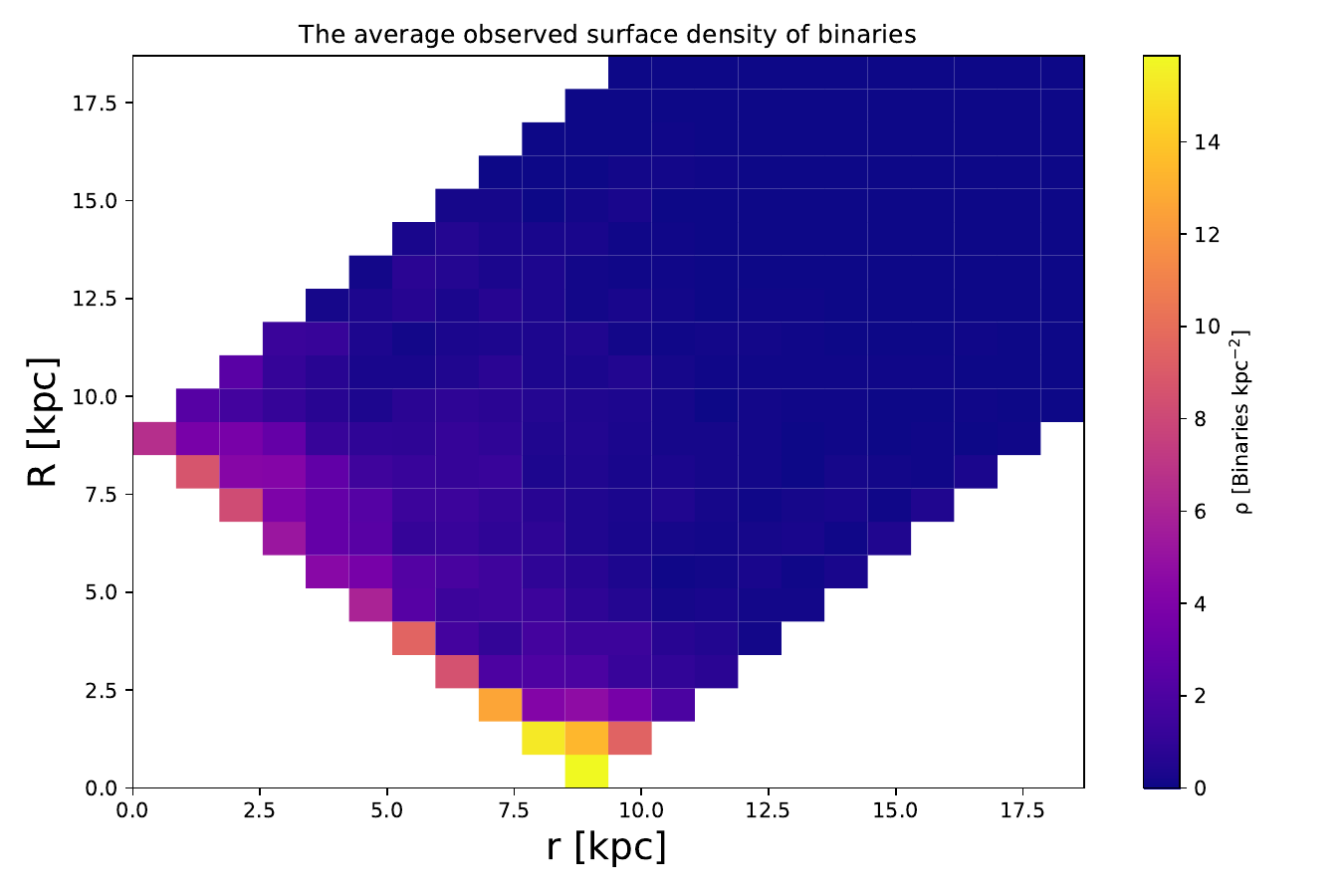}
\caption{The average observed surface density of {XRBs} across the Galactic plane. $\rm R$ is the distance from the Galactic Center and $\rm r$ is the distance from the Sun.}
\label{Table2D_HMXBs}
\end{figure}

To account for the decrease in binary density with increasing distance from the Sun, the radius of each grid cell increases with distance. Specifically, the radius $\rm R_{ci}$ of each grid cell is given by the following relation:
\begin{equation}
R_{ci} = 0.85 \times \left( 1.2\times(i - 1) / 22 + 1 \right) ~\mathrm{kpc},\notag
\end{equation}
where $\rm i$ ranges from 1 to 23, corresponding to different radial layers extending up to the maximum considered distance of 18.7 kpc. Larger grid cell radii at greater distances help mitigate density fluctuations due to small-number statistics. The apparent binary densities, in units of {binaries} per square kiloparsec ($\rm binaries ~kpc^{-2} $), for each grid cell (silver circles in Figure \ref{projection}) are presented in Figure \ref{Table2D_HMXBs}. Due to the stronger penetrating power of hard X-rays~\citep{2010exru.book.....S}, we have not considered the distance selection effect here.

\subsection{Distribution}
The spatial distribution of {XRBs} within the Milky Way is constructed by taking the weighted average of the observed {XRB} surface density (SD) shown in Figure \ref{Table2D_HMXBs}.

Due to the Poisson fluctuations in the number of {XRBs} ($\rm N_{ij}$) counted within each small circular spatial grid cell, it is necessary to account for the error in the observed SD, which can be calculated as follows ~\citep{yusifov2004revisiting, xie2024modeling}:
\begin{equation}
\frac{\sigma_{\rho_{i j}}}{\rho_{i j}} = \frac{\sqrt{N_{i j}}}{N_{i j}},
\end{equation}
where $\rm \sigma_{\rho_{i j}}$ is the error of $\rm \rho_{i j}$ in Figure \ref{Table2D_HMXBs}. We further obtain the radial distribution of {XRBs} by taking the weighted average as follows: 
\begin{equation}
\rho\left(R_i\right) = \frac{\sum_j \rho_{i j}\sigma_{i j}^{-2}}{\sum_j \sigma_{i j}^{-2}} \quad \text{and} \quad \sigma^{-2}\left(R_i\right) = \sum_j \sigma_{i j}^{-2}
\label{error}
\end{equation}
where $\rm i$ and $\rm j$ represent the position of ($\rm R_{i}$, $\rm r_{j}$), $\rm R_{i}=i\times0.85~kpc$ is the distance from the GC, and $\rm r_{j}=j\times0.85~kpc$ is the distance from the Sun.

\begin{figure}[t]
\centering
\includegraphics[width=\columnwidth]{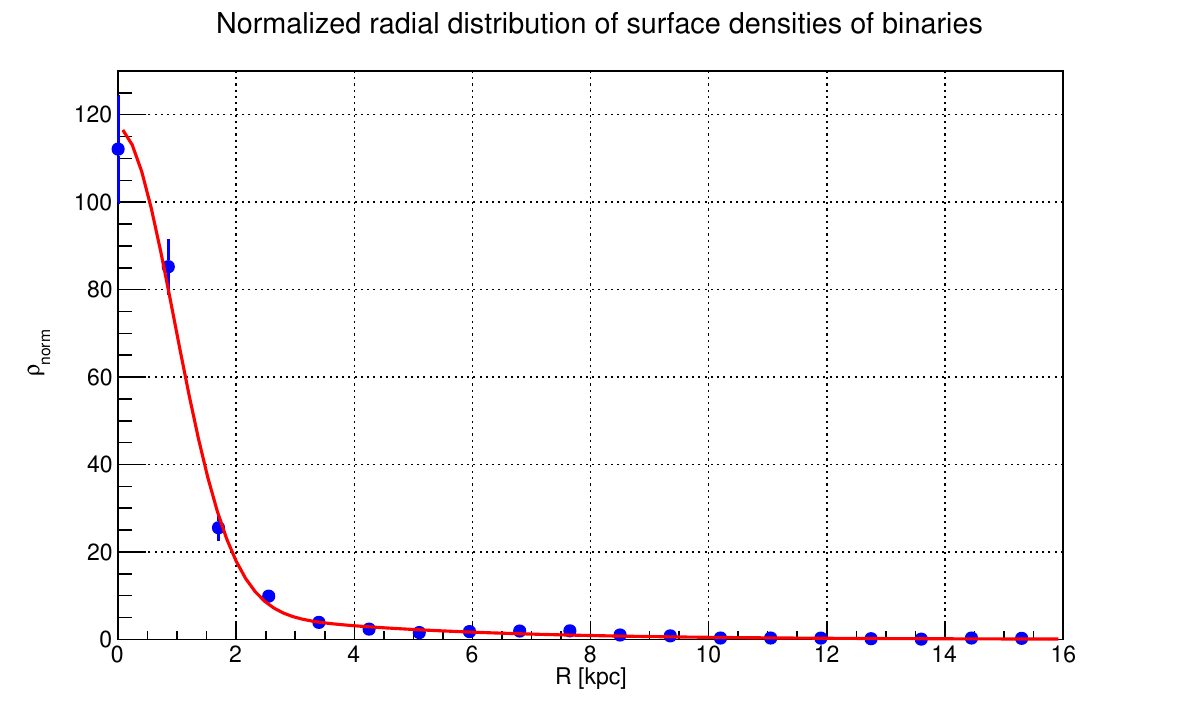}
\caption{Normalized radial distribution of surface density of {XRBs}. The error bars are derived from Equation \ref{error}. Here the data points (in blue) are fitted by the equation \ref{HMXBs_fit} (in red), and the chi-square is $\rm \chi^{2}/n.d.f={24.04/17}$. The data points and the fitted curve in the figure have been normalized to the value at the Sun's location.}
\label{HMXBs_distri}
\end{figure}

As shown in Figure \ref{HMXBs_distri}, the radial distribution of {XRBs} is fitted by 
\begin{equation}
{\rho(R)=A\left[\exp (-a\frac{R}{R_{\odot}})+B\exp (-b\frac{R^{2}}{R_{\odot}^{2}})\right]}
\label{HMXBs_fit}
\end{equation}
where $\rm R_{\odot}=8.5~kpc$ is the Sun-GC distance, {$\rm a =2.67\pm 0.21$, $\rm b =39.33 \pm 3.88$, $\rm A =1.54 \pm 0.29~XRBs~ kpc^{-2}$ and $\rm B =9.74 \pm 1.99$}. As a result, the source distribution of {XRBs} (or microquasars) normalized at the solar position can be written as: 
\begin{equation}
{H(R, z)=\frac{\exp (-a\frac{R}{R_{\odot}})+B\exp (-b\frac{R^{2}}{R_{\odot}^{2}})}{\exp (-a)+B\exp (-b)}\exp \left(-\frac{|z|}{z_s}\right)},
\label{HMXB_distri_norm}
\end{equation}
where $\rm z_s = 0.2~ kpc$. 

\section{Model and Methodology } \label{sec:method}
This paper considers X-ray binary star systems as the main contributors to the CR spectrum in the knee region in the whole Milky Way. However, the contribution in the low-energy range is still believed to primarily originate from SNRs.

After being injected into the vast interstellar medium, those accelerated cosmic-ray particles undergo a diffusion propagation process lasting millions of years within the turbulent magnetic field of the Galaxy. During this period, the cosmic-ray particles continuously interact with the interstellar medium, altering the characteristics of their energy spectrum. It is evident that Galactic cosmic rays primarily undergo two key stages: the primary acceleration process in the source region, and the long-term diffusion propagation process on Galactic scales. In fact, the propagation of cosmic rays is precisely described by the following equation~\citep{2007ARNPS..57..285S}:
\begin{equation}
\label{CRprop}
\begin{aligned}
\frac{\partial \psi(\vec{r}, p, t)}{\partial t}= & q(\vec{r}, p, t)+\vec{\nabla} \cdot\left(D_{xx} \vec{\nabla} \psi-\vec{V} \psi\right) \\
& +\frac{\partial}{\partial p} p^2 D_{pp} \frac{\partial}{\partial p} \frac{1}{p^2} \psi-\frac{\partial}{\partial p}\left[\dot{p} \psi-\frac{p}{3}(\vec{\nabla} \cdot \vec{V}) \psi\right] \\
& -\frac{1}{\tau_f} \psi-\frac{1}{\tau_r} \psi,
\end{aligned}
\end{equation}
where $\mathrm{\psi(}\vec{\rm r}\rm, p, t)$ represents the number density of cosmic rays with momentum $\rm p$ at position $\vec{\rm r}$, and $\mathrm{q(}\vec{\rm r}\rm, p, t)$ denotes the source term. $\rm D_{xx}$ is the spatial diffusion coefficient, $\vec{\rm V}$ is the convection velocity, $\rm D_{pp}$ connected to $\rm D_{xx}$ through this relation $\rm D_{pp}D_{xx}=\frac{4p^{2}v_{A}^{2}}{3\delta(4-\delta^2)(4-\delta)}$ is the diffusion coefficient in the momentum space, $\rm v_{A}=6~km/s$ is the Alfv\'{e}n speed, $\rm \delta$ is the slope of $\rm D_{xx}$, $\rm \tau_{f}$ is the timescale for fragmentation, and $\rm \tau_{r}$ is the timescale for radioactive decay. Therefore, this propagation equation encompasses sources (including the spatial distribution and injection spectrum of sources), diffusion-convection process, reacceleration process, energy loss process, spallation and decay process. Having undergone extensive scattering and prolonged diffusion and propagation, the cosmic rays that reach Earth have long attained a steady state, so that $\frac{\mathrm{\partial \psi(}\vec{\rm r}\rm, p, t)}{\rm \partial t}=0$.

The key aspect of this work is the assumption that low-energy ($\rm <100 ~TeV$) cosmic rays originate from Supernova Remnants (SNRs), while PeV cosmic rays originate from binary systems. However, the spatial distribution within the Galaxy and the injection spectra of these two types of sources differ.
Therefore, similar to previous work, we describe the spectrum injected by SNRs into the interstellar medium as:
\begin{equation}
Q(\mathcal{R})=q_0\mathcal{R}^{-\nu_1}\left[\left(1+\frac{\mathcal{R}} {\mathcal{R}_{br}}\right)^{\frac{(\nu_2-\nu_1)}{s}})\right]^{-s} e^{-\mathcal{R} / \mathcal{R}_c}
\end{equation}
Where $\rm \mathcal{R}$ is the particle rigidity, $\rm q_0$ is the normalization factor of the injection spectrum, $\rm \mathcal{R}_{br}$ is the break rigidity,The smoothness parameter $\rm s = 0.5$, $\rm \nu_1$ and $\rm \nu_2$ are the spectral indices before and after the break, $\rm \mathcal{R}_c$ is the cutoff rigidity, and $\rm e^{-\mathcal{R}/\mathcal{R}_c}$ describes the acceleration limit of SNRs. Consequently, the source term of Equation \ref{CRprop} for SNRs can be expressed as $\mathrm{q(}\vec{\rm r}\rm, p)=f(r,z)Q(\mathcal{R})$, where the source distribution $\rm f(r,z)$ is given by Equation \ref{SNRdistri}.

For PeV cosmic rays originating from binaries, taking SS 433 as an example ~\citep{2024arXiv241008988L}, its TeV radiation morphology suggests that radiation below several tens of TeV primarily originates from leptonic processes, while radiation above this energy range originates from the interaction of accelerated protons with molecular clouds near the central black hole. This implies that protons in the $\rm <100$ TeV energy range do not diffuse into the molecular cloud region; consequently, only the ultra-high-energy PeV cosmic rays can escape the source region and be injected into the interstellar medium. Microquasars, which contribute significantly to ultra-high-energy cosmic rays, as suggested by LHAASO observations and the theory of shock diffusion acceleration, exhibit an injection spectrum in the form of a truncated power-law spectrum:
\begin{equation}
Q(E)=Q_0(E / \mathrm{GeV})^s e^{-E / E_{\rm max}}
\label{XRBinj}
\end{equation}
with $\rm Q_0$ as the normalization factor, $\rm s$ as the spectral index, and $\rm E_{max}=2.5 ~PeV$. It should be specifically noted that due to the non-linear effects in the shock acceleration of ultra-high-energy cosmic rays, the spectral index $\rm s$ here is harder than the traditional value of -2.0 suggested by Fermi acceleration theory~\citep{1999ApJ...526..385B,2002APh....16..429B}. Therefore, we assume it to be -1.8 in this context. These values are derived from observational constraints on SS 433 by LHAASO, which also applies to other microquasars observed by LHAASO. {Here, SS 433 serves as an exemplary microquasar among those observed by LHAASO. The extended ultra-high-energy radiation from its central region has been detected by LHAASO to be spatially correlated with a giant gas cloud. This indicates that the ultra-high-energy radiation from SS 433 originates from hadronic processes~\citep{2024arXiv241008988L}.} Consequently, the source term for microquasars is $\mathrm{q(}\vec{\rm r}\rm, p)=H(r,z)Q(E)$.

In our model, for each type of source (SNR or binary), the cosmic-ray power injected into the interstellar medium per individual source and the production rate are constant. However, the source distribution is position-dependent within the Galaxy. Consequently, the cosmic-ray production rate varies across different Galactic regions due to differences in the number density of sources.

In our propagation framework, CR diffusion is spatially dependent, meaning the diffusion of CRs depends on the distribution of CR sources. In this paper, although we consider binary stars as the primary contributors to the CR spectrum in the knee region, we assume that the diffusion of CRs depends on the distribution of pulsar sources $\rm f(r, z)$, which is given by the Equation \ref{SNRdistri}. High-energy CRs accelerated by binary star systems and CRs accelerated by SNRs propagate in such an environment. The diffusion coefficient is described as~\citep{2016ApJ...819...54G,2018PhRvD..97f3008G}
\begin{equation}
D_{x x}(r, z, \mathcal{R})=D_0 F(r, z) \beta^\eta\left(\frac{\mathcal{R}}{\mathcal{R}_0}\right)^{\delta_0 F(r, z)}
\label{eq3.1}
\end{equation}
where the function $\rm F(r, z)$ is defined as:
\begin{equation}
    F(r, z)=\left\{ \begin{array}{ll}
        g(r, z)+[1-g(r, z)]\left(\frac{z}{\xi z_0}\right)^n, &  |z| \leq \xi z_0\\
        1, &|z|>\xi z_0 
    \end{array}\right.
\end{equation}
with $\rm g(r, z) = N_m/[1 + f(r, z)]$, $\rm N_{m}=0.62$, $\rm \xi=0.1$, $\rm n=4.0$ and the half height of the Galaxy $\rm z_{0}=5~kpc$. Here, the diffusion coefficient in the outer diffusive zone is in agreement with the conventional propagation model, whereas in the inner zone, it shows an anti-correlation with the source distribution $\rm f(r,z)$, given by~\citep{1998ApJ...504..761C}:
\begin{equation}
f(r, z)=\left(\frac{r}{r_{\odot}}\right)^{1.25} \exp \left[-\frac{3.87\left(r-r_{\odot}\right)}{r_{\odot}}\right] \exp \left(-\frac{|z|}{z_s}\right),
\label{SNRdistri}
\end{equation}
where $\rm r$ is the radial distance to the Galactic center, $\rm z$ represents the height from the Galactic plane, $\rm r_\odot = 8.5~ kpc$ and $\rm z_s = 0.2~ kpc$.

\subsection{Nearby Source}
The observed bubble structure in the CR proton spectrum between 200 GV and 14 TV, along with the anisotropic phase of CRs in this energy range pointing towards the anti-Galactic center, suggests the possible existence of a nearby source in the anti-Galactic center direction. To reproduce the observational data, we introduce a nearby source in this direction and consider its injection spectrum as an instantaneous injection, with the injection rate assumed to be,
\begin{equation}
Q(\mathcal{R}, t)=q_0 \delta\left(t-t_0\right)\left(\frac{\mathcal{R}}{\mathcal{R}_0}\right)^{-\gamma} \exp \left[-\frac{\mathcal{R}}{\mathcal{R}_{c}}\right]
\end{equation}
where $\rm \mathcal{R}$ represents the rigidity, $\rm \gamma$ is the spectral index, $\rm \mathcal{R}_c$ is the cutoff rigidity and $\rm t_0$ is the time of the supernova explosion. According to previous study, we assummed the local source located at ($\rm R.A.= 4^h0^m$, $\rm \delta = 24^{\circ} 30'$) with a distance of $\rm \sim 0.3 ~kpc$ an age of $\rm 3 \times 10^5 ~yr$, which is closed to the Geminga~\citep{2019JCAP...10..010L,2019JCAP...12..007Q}.  the normalization factor $\rm q_0$ is determined through fitting the energy spectra of CR protons.

\section{Results and discussion}\label{sec:results}
\subsection{$\rm B/C ratio$} \label{subsec:results_LHAASO}
\begin{figure}[t]
\centering
\includegraphics[width=\columnwidth]{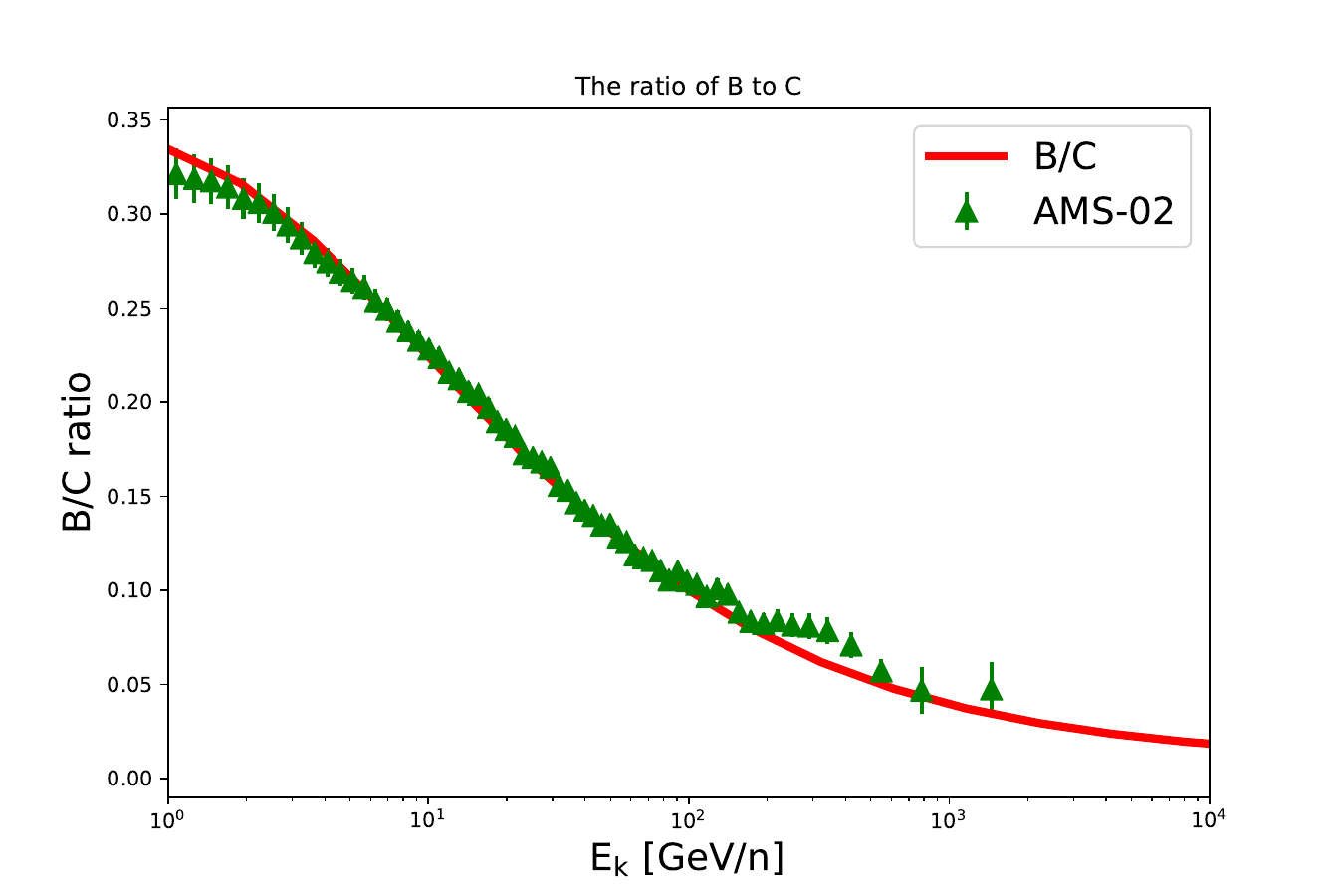}
\caption{The ratio of boron-to-carbon. The data points are taken from the AMS-02~\citep{2016PhRvL.117w1102A}. $\rm E_{k}$ refers to the kinetic energy per nucleon of cosmic rays. Cosmic rays with energies below 10 GeV are influenced by solar modulation, but our theoretical calculations do not take this solar modulation effect into account.}
\label{ratio}
\end{figure}

The B/C ratio can help us determine the diffusion coefficient. In the scenario of CRs propagating from their sources to the solar system through diffusion, the ratio of secondary CR particles to primary CR particles gives us detailed information of the diffusion process. As shown in Figure \ref{ratio}, there is good consistency between the result of the B/C ratio calculated by our model (in red) and the experimental observations of AMS-02 (in green). Thus, the rigidity-dependent diffusion coefficient parameterised by equation \ref{eq3.1} is completely obtained, where $\rm D_{0}=4.0\times 10^{24}~m^{2}s^{-1}$ and $\rm \delta_{0}=0.58$ are manually tuned in this work.

The hardening of the B/C energy spectrum around 200 GeV/n, which deviates from a single power-law distribution, is mainly attributed to propagation, acceleration or a local source. Observationally, secondary CR species exhibit spectral hardening in the same energy range, and they are harder than the spectra of their primary counterparts~\citep{2018PhRvL.120b1101A}. This provides significant support for the origin of propagation. However, others suggest that this characteristic could originate from the contribution of a nearby source~\citep{2023FrPhy..1844301M}.

\subsection{The proton spectrum and CR anisotropy}
\label{subsec:proton_spec}
Since the opinion that SNRs accelerate cosmic rays to the PeV energies remains debated, XRBs (or microquasars) as one of the ``PeVatrons" should play an import role in the proton energy spectrum and CR anisotropy around the knee structure. In this scenario, both the total proton spectrum and CR anisotropy are contributed by the SNRs and the nearby source at low energies, and XRBs at high energies.

\begin{figure}[t]
\centering
\includegraphics[width=\columnwidth]{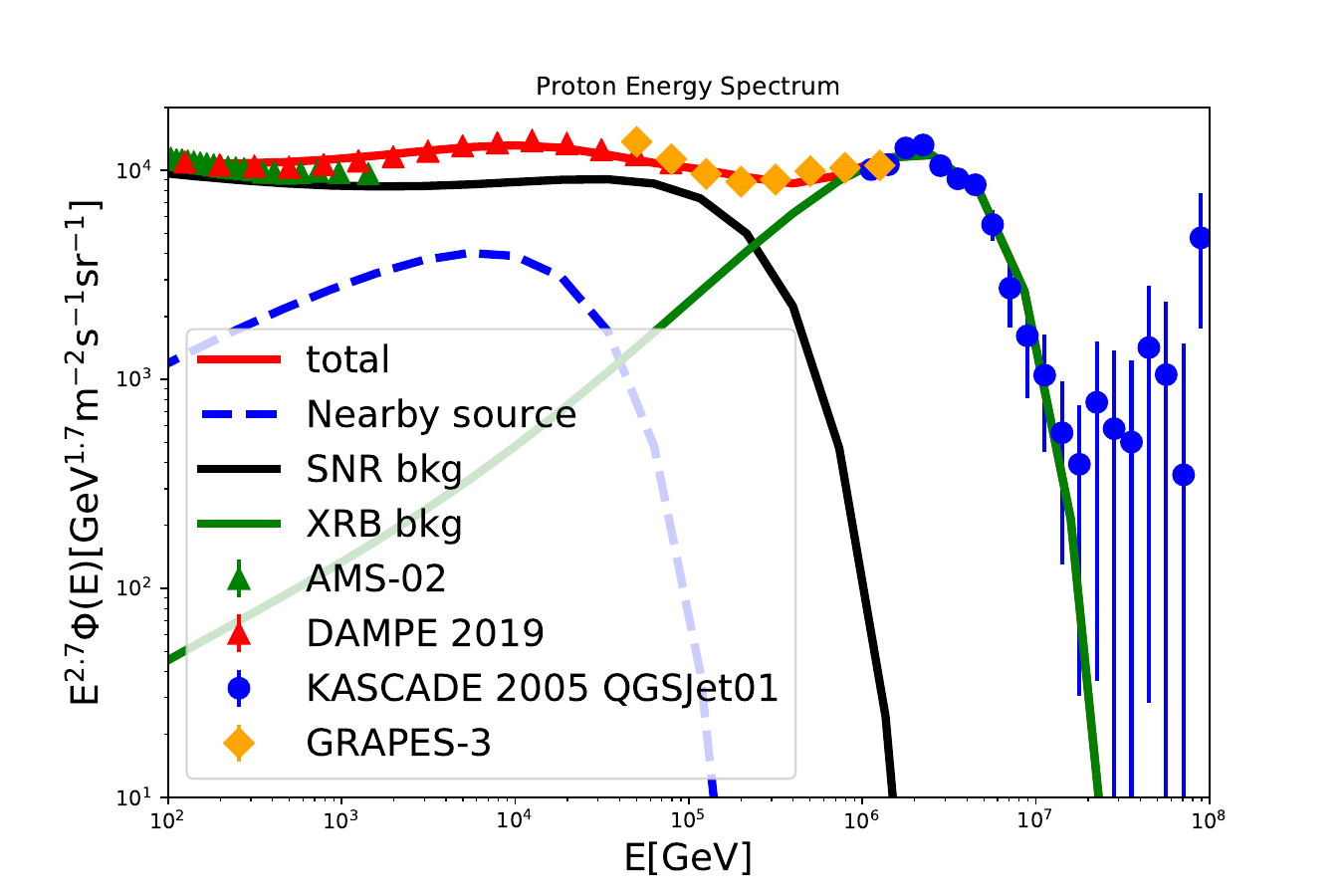}
\caption{ The calculated proton spectrum compared with observed data. The black solid lines is the fluxes from background SNRs, the green solid line is the contribution from the XRBs, and the blue dashed line is the contribution from the local SNR. The data points are taken from the AMS-02 (green)~\citep{2015PhRvL.114q1103A}, DAMPE (red)~\citep{2019SciA....5.3793A}, GRAPES~\citep{2024PhRvL.132e1002V}, KASCADE~\citep{2005APh....24....1A}.}
\label{SpecP}
\end{figure}

As a result, the calculated proton energy specrum by our model has shown good consistency with the observed proton energy spectra. As seen in Figure \ref{SpecP}, the total proton energy spectrum (in red) is the sum of the contributions of the SNRs, with a cutoff rigidity at 200 TeV (in black), the nearby source (in blue), and XRBs (or microquasars), with a cutoff rigidity at 2.5 PeV (in green). Here the contributions of SNRs and XRBs with different spatial distributions are calculated by DRAGON~\citep{2008JCAP...10..018E} (Diffusion of cosmic RAys in galaxy modelizatiON, a numerical code used to resolve the cosmic ray propagation equation) using the same spatially-dependent propagation (namely SDP) model. The result is perfectly consistent with experimental observations. However, it is worth noting that the measurements from KASCADE have large errors and fluctuations at high energies, primarily due to the limited number of measured showers~\citep{2005APh....24....1A}.

We further estimate the power of XRBs and SNRs. Considering that the total energy of SNRs is commonly known as $\rm E_{SNR}\approx10^{51}~erg$, the rate of SN explosions $\rm R_{SNR}\approx1-3~times~per~century$, the time of the acceleration process $\rm T_{acc}=1000 ~yrs$ and the efficiency $\rm \eta=10\%$, we calculate the power of accelerating CRs through SNRs $\rm P_{SNR,CR}=E_{acc}/T_{acc}=E_{SNR}R_{SNR}\eta\approx10^{41}~erg~s^{-1}$. For microquasars, we take SS 433~\citep{2024arXiv241008988L} as an example, the authors obtained the proton luminosity of SS 433 to be $\rm \sim 10^{38} ~erg ~s^{-1}$ and expected that the total proton injection luminosity from all Galactic microquasars may exceed $\sim$10 times that of SS 433. We adopt $\rm \sim 10^{39} ~erg ~s^{-1}$ as a representative value of microquasars.

\begin{figure}[ht!]
\centering
\includegraphics[width=0.9\columnwidth]{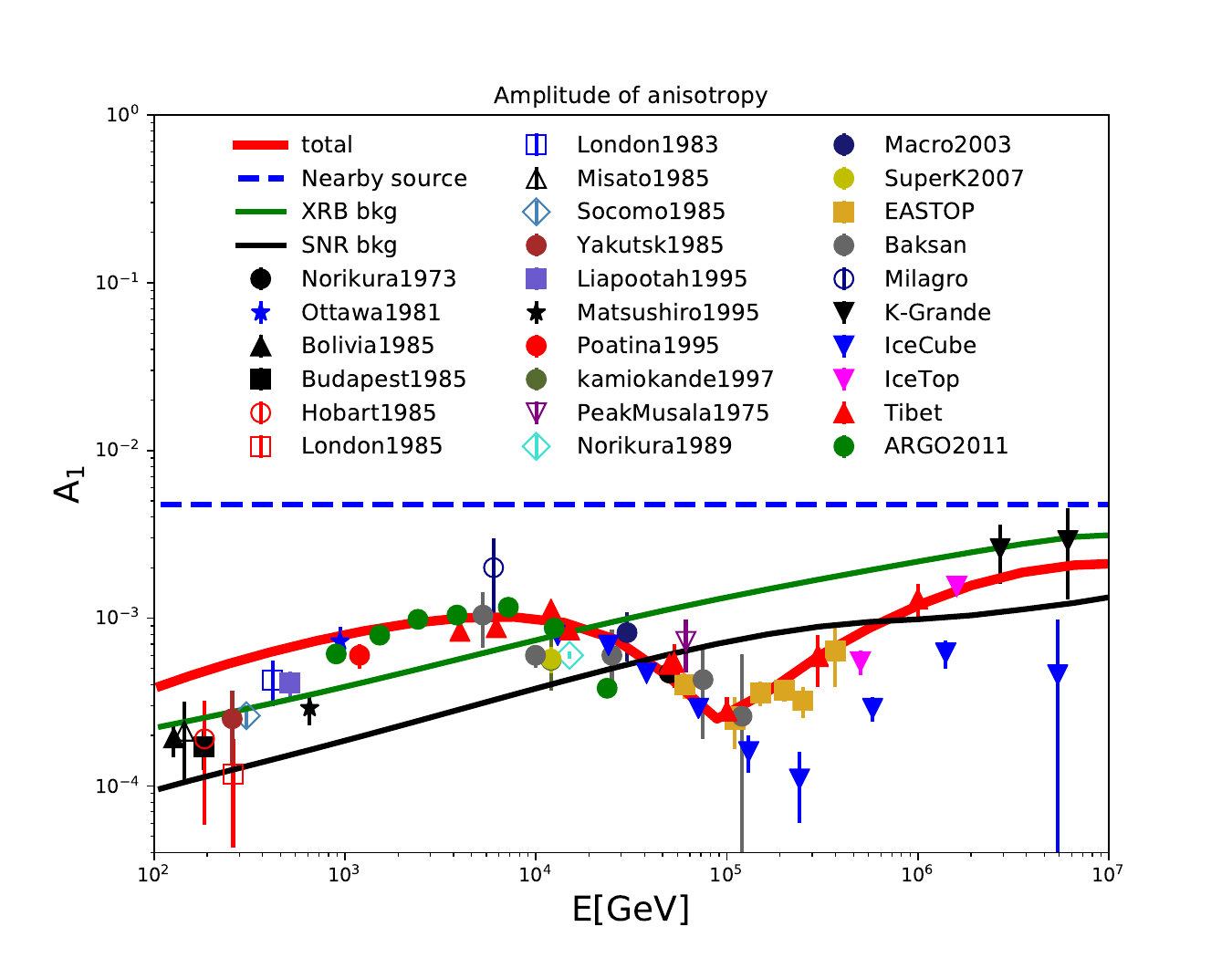}
\includegraphics[width=0.9\columnwidth]{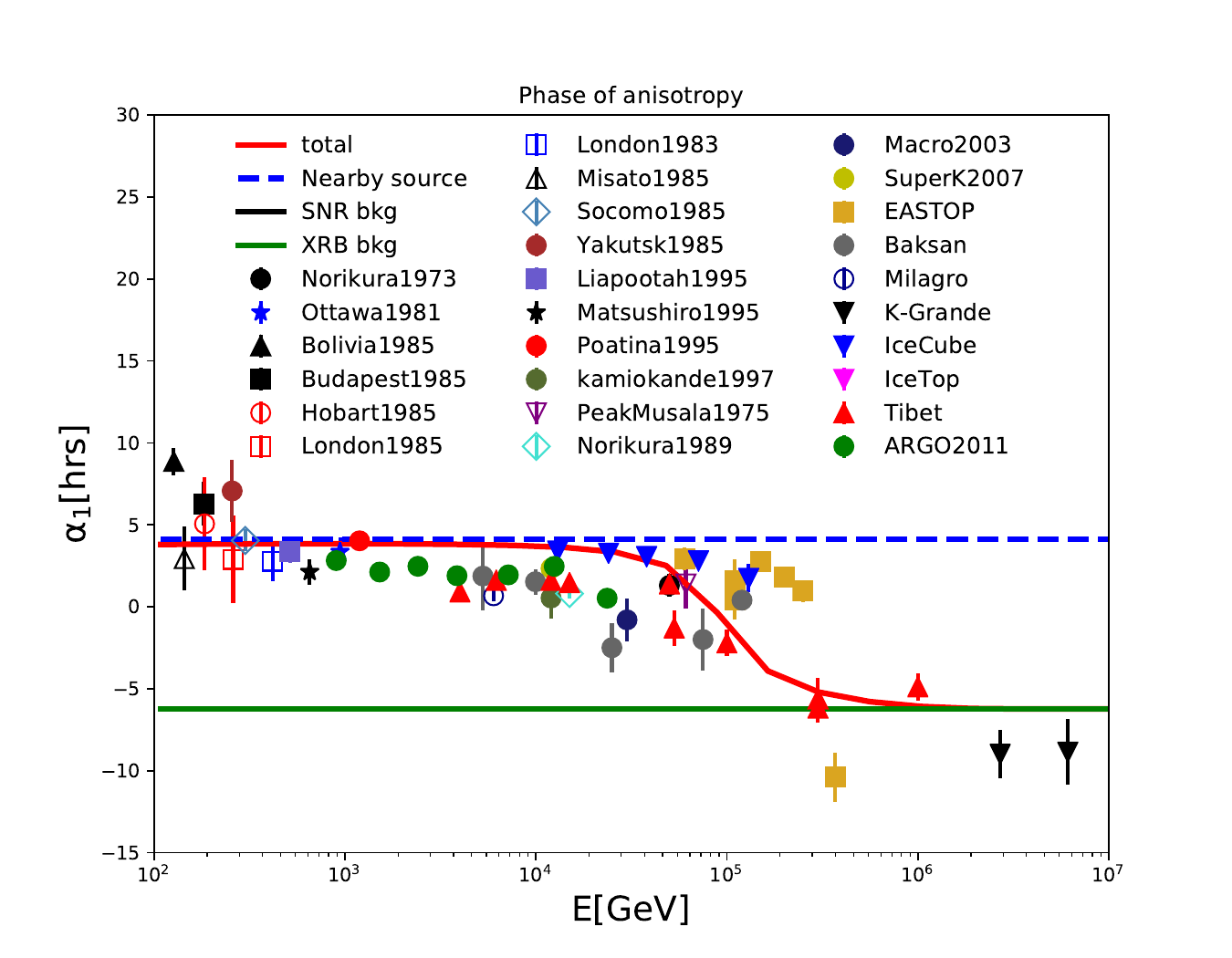}
\caption{The amplitude (top panel) and phase (bottom panel) of anisotropy of CRs. Here, the total amplitude of anisotropy (represented by the solid red line) is contributed by the SNRs, the X-ray binaries, and the nearby source, as is the case with the total phase of CR anisotropy. The data points are taken from  Norikura ~\citep{1973ICRC....2.1058S}, Ottawa ~\citep{1981ICRC...10..246B}, London ~\citep{1983ICRC....3..383T}, Bolivia ~\citep{1985P&SS...33.1069S}, Budapest
 ~\citep{1985P&SS...33.1069S}, Hobart ~\citep{1985P&SS...33.1069S}, London ~\citep{1985P&SS...33.1069S}, Misato ~\citep{1985P&SS...33.1069S}, Socorro ~\citep{1985P&SS...33.1069S}, Yakutsk ~\citep{1985P&SS...33.1069S}, Banksan ~\citep{1987ICRC....2...22A}, Hong Kong ~\citep{1987ICRC....2...18L}, Sakashita ~\citep{1990ICRC....6..361U}, Utah ~\citep{1991ApJ...376..322C},
 Liapootah ~\citep{1995ICRC....4..639M}, Matsushiro ~\citep{1995ICRC....4..648M}, Poatina ~\citep{1995ICRC....4..635F}, Kamiokande ~\citep{1997PhRvD..56...23M},
 Marco ~\citep{2003PhRvD..67d2002A}, SuperKamiokande ~\citep{2007PhRvD..75f2003G}, PeakMusala
 ~\citep{1975ICRC....2..586G}, Baksan ~\citep{1981ICRC....2..146A}, Norikura ~\citep{1989NCimC..12..695N}, EAS-TOP ~\citep{1995ICRC....2..800A,1996ApJ...470..501A,2009ApJ...692L.130A},
 Baksan ~\citep{2009NuPhS.196..179A}, Milagro ~\citep{2009ApJ...698.2121A}, KASCADE-Grande ~\citep{2015ICRC...34..281C},IceCube ~\citep{2010ApJ...718L.194A,2012ApJ...746...33A}, Ice-Top ~\citep{2013ApJ...765...55A}, ARGO
YBJ ~\citep{2015ApJ...809...90B}, Tibet ~\citep{2005ApJ...626L..29A,2017ApJ...836..153A,2015ICRC...34..355A}, AUGER ~\citep{Aab_2020}.}
\label{Anisotropy}
\end{figure}

On the other hand, the observed energy dependence of CR anisotropy is also described well by our model. The total amplitude (in red, top panel) and phase (in red, bottom panel) of anisotropy in Figure \ref{Anisotropy} calculated by our model fitting well with experimental observations have no extra features around the knee structure. 

The large scatter among the data of anisotropy observations of different experiments has sparked some discussion. Some explanations are proposed~\citep{2017PrPNP..94..184A} that the one-dimensional anisotropy from the right-ascension projection, observed from anisotropic experiments, are influenced by the limited field of view in each of these experiments, necessitating appropriate geometric corrections, especially the early experimental observations, which causes the different experimental observations to be too dispersed from each other. The latest anisotropy observations of IceCube~\citep{2025ApJ...981..182A} have adopted this opinion.

\subsection{Diffuse gamma ray emission}
\label{subsec:diffuse}
\begin{figure}[ht!]
\centering
\includegraphics[width=\columnwidth]{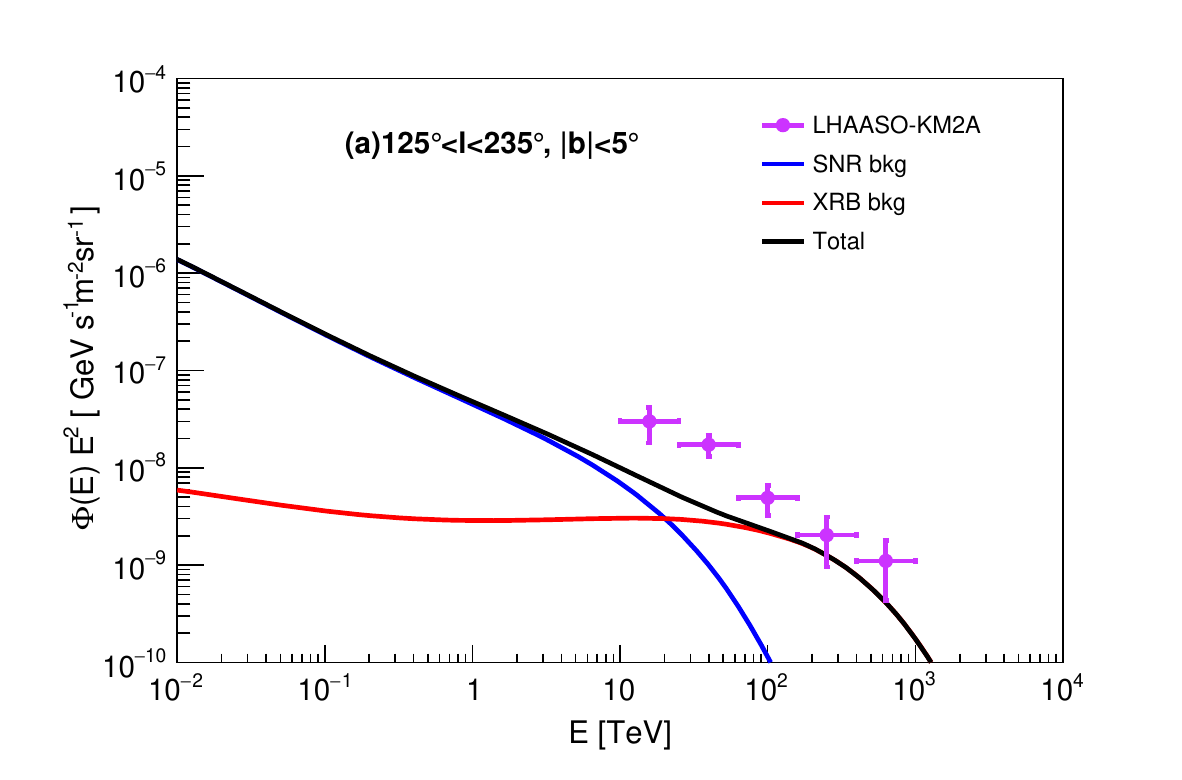}
\includegraphics[width=\columnwidth]{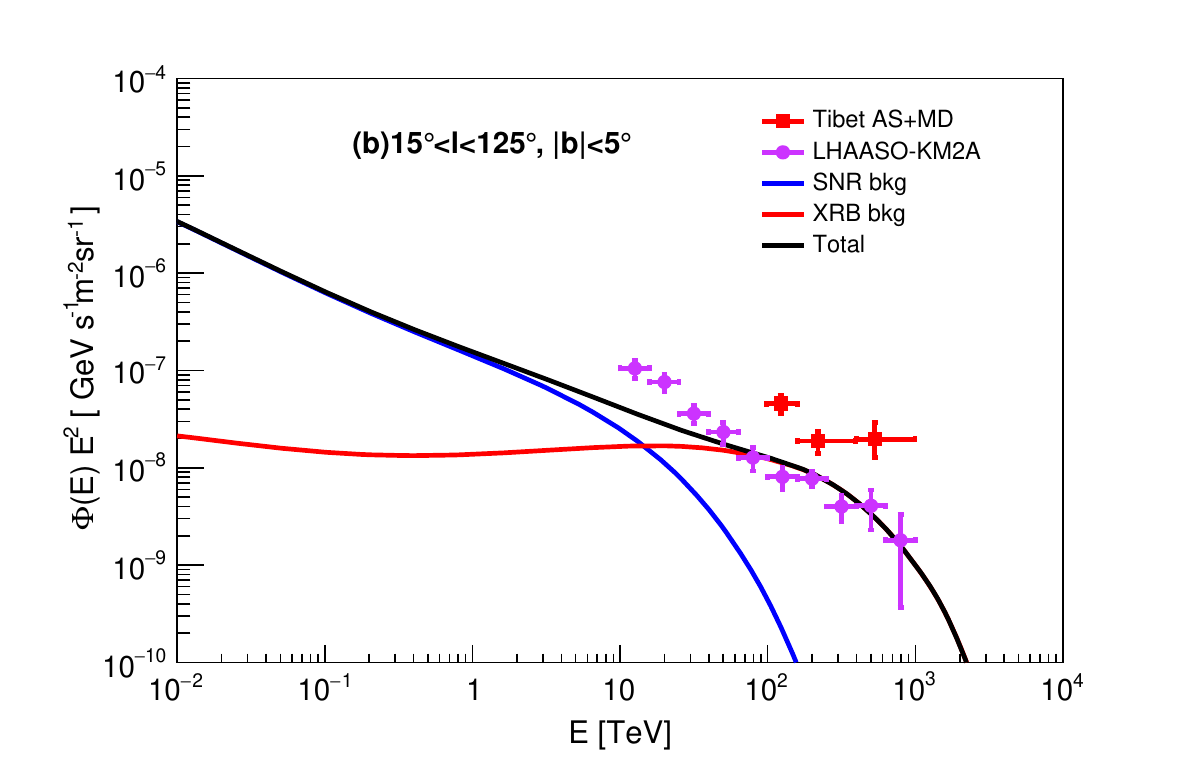}
\caption{Diffuse gamma energy spectra in the Galactic disk. The top panel: with Galactic longitude between $\rm 125^{\circ}$ to $\rm 
235^{\circ}$; the bottom panel: with Galactic longitude between $\rm 15^{\circ}$ to $\rm 
125^{\circ}$. The solid black line predicted by our model represents the total diffuse gamma, which is the sum of diffuse gamma contributed by SNRs (the solid blue line) and X-ray binaries (the solid red line). In addition, the absorb effect has already been taken into consideration in our model. The data points are taken from Tibet AS+MD~\citep{2021PhRvL.126n1101A} and LHAASO-KM2A~\citep{2023PhRvL.131o1001C}.}
\label{Gamma}
\end{figure}

CR particles escape from their sources and diffuse into the interstellar medium, filling the entire space of the Milky Way galaxy. During their propagation, they interact with the surrounding gas through proton-proton process, generating diffuse gamma rays. Therefore, diffuse gamma rays can serve as probes to reflect the spatial distribution characteristics of CRs. Both the Tibet-$\rm AS\gamma$ and the LHAASO experiments have detected diffuse gamma rays with energies above 100 TeV, which may result from CR particles, accelerated to PeV energies in certain PeVatrons, such as some XRBs, interacting with the interstellar medium and producing diffuse gamma radiation. In this work, we consider that the CR particles at the PeV energy range are primarily accelerated by XRBs. By reproducing the CR proton spectrum and anisotropy observational data as constraints, we calculate the diffuse gamma-ray radiation produced by this group of CR protons. As shown in Fig.\ref{Gamma}, we can seem to infer that although the expected flux from the model is lower than the observed data, XRBs may be the dominant contributor to the sub-PeV CRs.

The two panels in the figure \ref{Gamma} show a significant deviation of the model from the observed data at around 10 TeV.
According to our model's framework, low-energy cosmic rays are collectively contributed by distant sources throughout the Galaxy, while high-energy cosmic rays are primarily dominated by local sources. Although the analysis of the diffuse gamma-ray emission observed by LHAASO masked known point sources and extended sources detected by KM2A and other experiments, residual contamination from some extended and point sources was found to significantly contribute to the diffuse gamma-ray radiation ~\citep{2024ApJ...974..276N,2024ChPhC..48k5105C}. Consequently, the diffuse gamma-ray spectrum predicted by our model from cosmic rays is lower than the experimental observational data. Additionally, the Tibet AS$\gamma$ observational data shows inconsistency with LHAASO data. This discrepancy may arise from different source subtraction procedures applied when analyzing the Tibet AS$\gamma$ and LHAASO datasets, leading to variations in the covered Galactic latitude range between the two experiments ~\citep{2024ApJ...977L...3K,2021PhRvL.126n1101A,2023PhRvL.131o1001C}. Therefore, On one hand, LHAASO may mask a significant portion of the sky to remove contamination from resolved gamma-ray sources, potentially leading to an underestimation of the diffuse gamma-ray flux ~\citep{2023ApJ...957L...6F}. On the other hand, the diffuse flux measured by Tibet AS$\gamma$ could be substantially contaminated by emissions from Galactic gamma-ray sources due to its less comprehensive source-masking scheme ~\citep{2024NatAs...8..628Y}.

\section{Summary and Conclusion} \label{sec:summary}
Although it is theoretically believed that CRs in the knee region and below originate from the Milky Way, the exact type of astrophysical sources that dominate the CRs near the knee region remains an unresolved issue. LHAASO, with its one-square-kilometer detection area and powerful background noise suppression capabilities, has observed PeV radiation from pulsars and XRBs, revealing their potential role as major contributors to CRs in the knee region.

In this work, we take the detection of PeV (Peta-electronvolt) radiation from microquasars by LHAASO as our basis and consider X-ray binaries as the primary contributors to the ``knee" region of CRs. By statistically analyzing the distribution of currently observed XRBs in the Galactic disk, we assume that they continuously inject CR protons into the Milky Way in a spectral form similar to the source SS443, and provide a non-negligible contribution to the CR distribution throughout the entire Galaxy. 

We have analyzed the distribution of sources contributing to CRs. Based on this distribution, we have effectively modeled the knee region. We used the DRAGON to calculate the contribution of microquasars to the knee of the high energy CRs for the source distribution (Equation \ref{HMXB_distri_norm}) and the injection spectrum (Equation \ref{XRBinj}). Our calculations show that microquasars can make a significant contribution to the CRs in the ``knee" region. On the other hand, considering microquasars as contributors to the Galactic CRs also provides a good fit to the observed anisotropy of CRs and the diffuse gamma-ray emission from the Galactic disk observed by LHAASO. Theoretically, although considering SNRs as acceleration sites for ultra-high-energy CRs, and using a space-dependent model, can also reproduce the CR energy spectrum, anisotropy ~\citep{2023ApJ...956...75Q,2024PhRvD.109f3001Y,2024ApJ...974..276N}, and diffuse gamma-ray emission ~\citep{2024ApJ...964...28H,2018PhRvD..97f3008G} from the Galactic disk in the PeV energy range, there is, to date, almost no observational evidence that SNRs can accelerate CRs to the PeV energy range. In contrast, more and more microquasars observed by LHAASO were found to emit photons with energies exceeding 100 TeV. This is almost identical to our results, which suggest that microquasars are likely the main contributors to Galactic ``knee" CRs.

{It is noteworthy that in our study, the spatial distribution of XRBs is normalized to its value at the solar position. Consequently, the energy spectrum and anisotropy at the solar position remain insensitive to the spatial distribution of sources away from the solar position. Furthermore, as shown in Figure \ref{Gamma}, the diffuse gamma rays currently observed by LHAASO are large-scale and not aligned with the Galactic center direction, thereby demonstrating insensitivity to the distribution of XRBs. In addition, the assumption that cosmic-ray diffusion depends on the source distribution of SNRs leads to the B/C ratio being unaffected by the spatial distribution of XRBs.}

With the increasing number of observed microquasars in binary systems, the methodology presented in this study will enable independent and precise quantification of their spatial distribution. Consequently, the contribution of microquasars to the cosmic-ray knee region will become increasingly significant.

In conclusion, the combination of observational data and theoretical modeling has provided significant insights into the complex mechanisms governing CR propagation, and further discoveries in the PeV energy range will continue to enhance our understanding of this phenomenon.

\section*{acknowledgments} 
This work is supported in China by the National Key R$\&$D program of China under the grant 2024YFA1611402 and the Natural Sciences Foundation of China (No. 12333006, 12375108 and 12275279).

\bibliography{sample631}{}
\bibliographystyle{aasjournal}

\end{document}